\documentclass[prb,twocolumn,superscriptaddress,nofootinbib]{revtex4-2}

\usepackage{amsmath}
\usepackage{graphicx}
\usepackage{xcolor}
\usepackage{hyperref}

\usepackage{amsmath,amssymb}
\newcommand{\tr}[1]{\textcolor{blue}{#1}}

\begin{document}

\title{Phase Diagram and Ashkin–Teller Universality in the Classical Square-Lattice Heisenberg–compass Model}

\author{Yuchen Fan}
\email{yuchenfan@imu.edu.cn}
\affiliation{Research Center for Quantum Physics and Technologies, School of Physical Science and Technology, Inner Mongolia University, Hohhot 010021,China}

\begin{abstract}
We determine the finite-temperature phase diagram and critical behavior of the classical square-lattice Heisenberg--compass model using large-scale Monte Carlo simulations and finite-size scaling. Six symmetry-distinct ordered phases are identified. The four phases that simultaneously break the spin--lattice $C_4$ and in-plane spin-inversion symmetries undergo continuous transitions in the Ashkin--Teller universality class, with the associated critical lines terminating at four-state Potts points, beyond which the transitions become first order. In contrast, the two $z$-polarized phases display conventional two-dimensional Ising criticality. Our results reveal how the interplay between Heisenberg exchange and compass anisotropy organizes these distinct critical regimes, thereby completing the characterization of the model’s thermal phase transitions.
\end{abstract}

\maketitle

\section{Introduction}
The interplay of isotropic and anisotropic spin interactions is a central theme in frustrated magnetism. 
In particular, compass-type interactions, which couple spin components depending on the spatial orientation of the bond, arise naturally in spin--orbit--coupled Mott insulators and related materials, and have been extensively investigated as platforms for exotic phases ranging from nematic order to spin liquids~\cite{RevModPhys.87.1,PhysRevB.72.024448,PhysRevLett.102.017205,KITAEV20062,RevModPhys.89.025003,Savary2017}. Among these, the Heisenberg–compass model on the square lattice provides a minimal setting in which Heisenberg exchange competes with bond-directional compass interactions, giving rise to a rich landscape of accidental degeneracies and fluctuation-induced order selection~\cite{Trousselet_2010,PhysRevB.86.134412,Vladimirov2015,PIRES2018326,PhysRevLett.130.266702,PhysRevB.110.104426}.

Recent efforts have established key aspects of the \textit{classical} Heisenberg–compass model at zero and low temperatures~\cite{PhysRevLett.130.266702,PhysRevB.110.104426}. 
Khatua \textit{et al.} showed that, in the ferromagnetic regime, thermal fluctuations generate a pseudo-Goldstone gap with a universal square-root temperature dependence, providing a clear dynamical signature of order-by-thermal-disorder~\cite{PhysRevLett.130.266702}. 
In a subsequent work, they mapped out the classical ground-state manifold of the Heisenberg--compass model, identifying six ordered phases and elucidating the roles of order-by-disorder and Klein duality in shaping the overall phase diagram~\cite{PhysRevB.110.104426}. 
They also proposed a tentative finite-temperature phase diagram based on peaks in the specific heat~\cite{PhysRevB.110.104426}. 
While these studies have significantly advanced our understanding of the ground states and low-temperature fluctuations, the nature and universality of the thermal phase transitions remain unresolved.

In parallel, significant progress has been made in understanding an XY-type system with direction-dependent interactions—the ``generic compass model'' (gCM)~\cite{PhysRevB.109.195131}. 
Zhang, Guo, and Kaul showed that the gCM realizes Ashkin--Teller (AT) criticality: a line of continuous transitions with continuously varying exponents, terminating at a four-state Potts point and connecting to a line of first-order transitions (for background on AT criticality, see Refs.~\cite{Baxter1982,Nienhuis1987,DELFINO2004521}).
In the isotropic ($O(2)$) limit, the gCM undergoes a Kosterlitz--Thouless transition, while previous studies have so far been restricted to its ferromagnetic sector~\cite{PhysRevB.109.195131}. 
The AT universality in the gCM arises from the simultaneous breaking of spin-inversion and spin--lattice $C_4$ symmetries, occurring without an intermediate nematic regime. 
Whether the Heisenberg--compass model exhibits the same locked symmetry breaking or hosts a distinct nematic phase remains an open question.
Similar AT criticality has also been observed in a planar-rotator model with explicit fourfold anisotropy~\cite{PhysRevB.69.174407}—reducing $O(2)$ to $C_4$—and in frustrated spin systems such as the square-lattice $J_1$--$J_2$ Ising model~\cite{PhysRevLett.108.045702,PhysRevB.87.144406}, the $J$--$Q_3$ model~\cite{PhysRevB.87.180404}, and the frustrated bilayer Heisenberg model~\cite{Fan2024}.

\begin{figure}
\includegraphics[height=9.0cm,width=8.6cm]{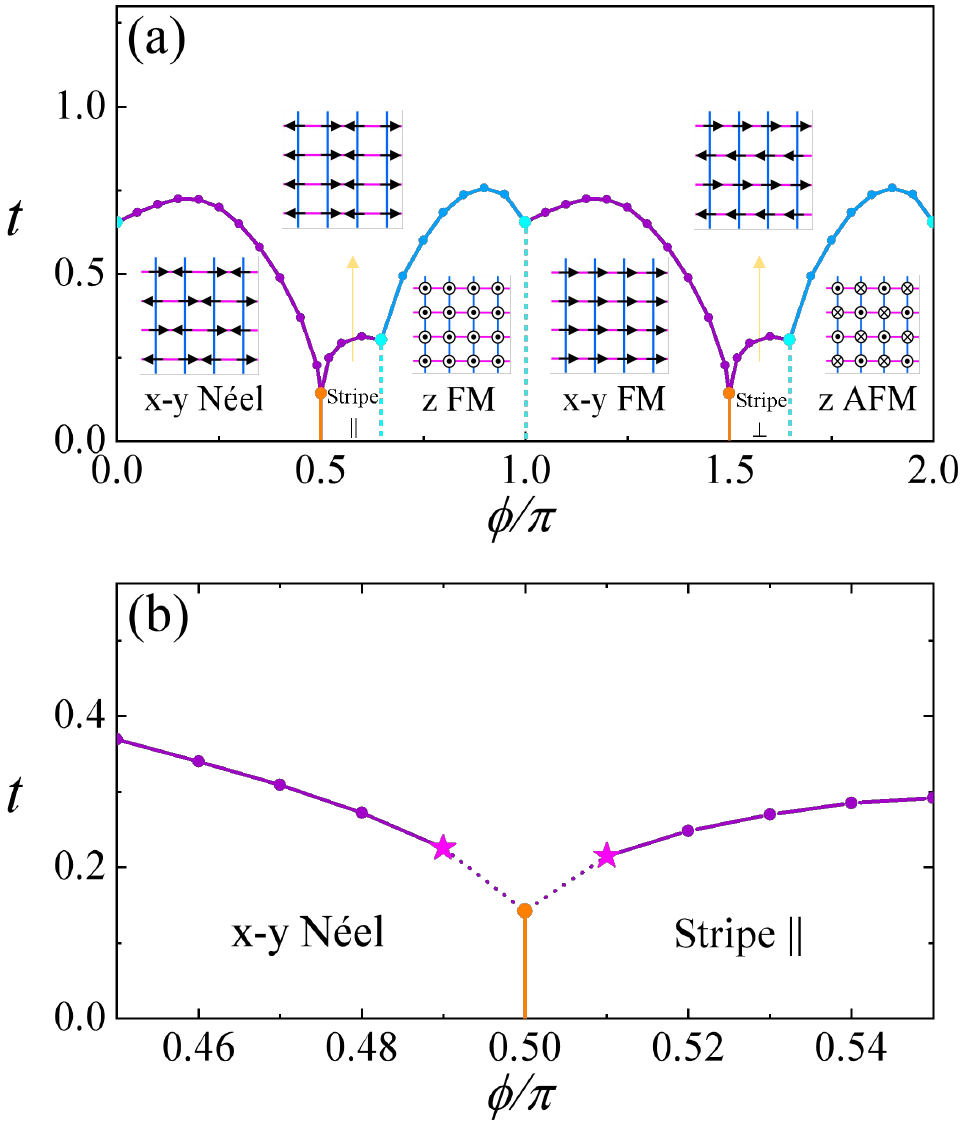}
\caption{
(a) Finite-temperature phase diagram of the classical Heisenberg--compass model in the $(t,\phi)$ plane, showing six ordered phases: x-y N\'eel, Stripe $\parallel$, z FM, x-y FM, Stripe $\perp$, and z AFM.
Stripe$\parallel$ (Stripe$\perp$) denotes stripe states with spins aligned parallel (perpendicular) to the stripe direction. 
The low-temperature order parameters as functions of $\phi$ are summarized in Appendix~\ref{app:order}. 
Cyan dots mark points where only thermal crossovers occur, and the corresponding crossover temperatures are extracted from the broad specific-heat maximum~\cite{Deng2025}. In particular, the cyan dots at $\phi=0$ and $\phi=\pi$ correspond to explicitly $O(3)$-symmetric limits, while those at $\phi=\pi-\tan^{-1}(2)$ and $\phi=2\pi-\tan^{-1}(2)$ possess a hidden $O(3)$ symmetry under Klein duality~\cite{PhysRevB.110.104426}.
As the $O(3)$-symmetric limits are approached, the anisotropy-driven transition temperature is expected to vanish asymptotically ($t_c \to 0$) in the thermodynamic limit.
Dashed cyan lines only serve as guides to the crossover trajectories. At $\phi=\pi/2$ and $3\pi/2$ (orange), the model reduces to the pure compass limit with an Ising-type transition~\cite{PhysRevLett.93.207201}.
(b) Enlarged view near $\phi\!\approx\!\pi/2$: solid dark magenta lines denote continuous Ashkin--Teller transitions (with a $\phi$-dependent exponent $\nu$), the bright magenta stars mark the four-state Potts points, and dashed dark magenta lines indicate first-order transitions.
}
\label{fig:phase_diagram}
\end{figure}

In this work, we determine the finite-temperature phase diagram and reveal the associated universality classes of the classical Heisenberg--compass model on the square lattice. Using large-scale Monte Carlo simulations and finite-size scaling, we identify six ordered phases [Fig.~\ref{fig:phase_diagram}]. As shown in Fig.~\ref{fig:phase_diagram}(a), four phases simultaneously break the coupled spin--lattice $C_4$ symmetry and the in-plane spin-inversion symmetry, exhibiting AT criticality, while the two $z$-polarized phases undergo conventional Ising transitions. These AT lines terminate at four-state Potts points, connecting to first-order boundaries beyond them, as highlighted in Fig.~\ref{fig:phase_diagram}(b). At special symmetry points, an enhanced $O(3)$ symmetry leads to thermal crossovers, whereas the model reduces to known Ising transitions in the pure compass limits~\cite{RevModPhys.87.1,PhysRevLett.93.207201,PhysRevE.81.066702}. Although our results share the common motif of AT criticality with the gCM~\cite{PhysRevB.109.195131}, the Heisenberg–compass model exhibits a substantially richer phase structure. The additional spin degree of freedom stabilizes two distinct $z$-polarized phases and yields a complete set of six ordered phases—features absent in the two-component case. Our work thus fully clarifies the finite-temperature phase transitions of the model, offering insights into related spin–orbit–coupled magnetic systems~\cite{PhysRevLett.110.117207,PhysRevX.4.021051,PhysRevB.94.161118,annurev:/content/journals/10.1146/annurev-conmatphys-031218-013113,PhysRevB.83.155118}.

The rest of this paper is organized as follows. In Sec.~\ref{sec:model}, we introduce the classical Heisenberg--compass model on the square lattice,
outline its symmetries and ordered phases, and describe the Monte Carlo methods and physical observables used in our simulations.
Section~\ref{sec:results} presents our main numerical findings, establishing the AT character of the phase transitions through the analysis of the continuous transition regime, the four-state Potts point, and adjacent first-order boundaries.
Section~\ref{sec:discussion} concludes with a discussion and outlook.

\section{Model and method \label{sec:model}}
We consider the classical Heisenberg--compass model on the square lattice, defined by the Hamiltonian
\begin{equation}
\mathcal{H} = J \sum_{\langle ij \rangle} \mathbf{S}_i \cdot \mathbf{S}_j 
 + K \sum_{i} \left( S_i^x S_{i+\hat{x}}^x + S_i^y S_{i+\hat{y}}^y \right),
\label{eq:Hamiltonian}
\end{equation}
where $\mathbf{S}_i=(S_i^x,S_i^y,S_i^z)$ is a three-component classical unit vector at site~$i$. 
The first term represents the nearest-neighbor isotropic Heisenberg exchange of strength~$J$, while the second introduces bond-directional compass interactions of strength~$K$, coupling the $x$ and $y$ spin components along their respective lattice directions. 
Throughout this work we parameterize the couplings as 
$J = \cos\phi$ and $K = \sin\phi$, 
so that the relative strength and sign of the two interactions can be continuously tuned by the angle~$\phi$.
Varying~$\phi$ thus interpolates smoothly between the Heisenberg and pure compass limits. 

This Hamiltonian possesses several discrete symmetries. 
First, it is invariant under global sign flips of individual spin components, 
$S_i^\alpha \!\to\! -S_i^\alpha$ for $\alpha \in \{x,y,z\}$ and all sites $i$, 
whose combined operation corresponds to full spin inversion, $\mathbf{S}_i \!\to\! -\mathbf{S}_i$.  
Second, while the Heisenberg term alone possesses continuous $O(3)$ spin-rotation symmetry, 
the compass term reduces it to a discrete subgroup that couples spin and lattice rotations. 
Specifically, the Hamiltonian remains invariant under a joint $\pi/2$ rotation of both the square lattice and the in-plane spin components about the $S^z$ axis, $(S^x,S^y)\!\to\!(S^y,-S^x)$, defining a coupled spin–lattice $C_4$ symmetry. 
In addition, the model respects a Klein duality that connects distinct ordered phases~\cite{PhysRevB.110.104426}, 
as well as a sublattice spin-flip transformation that maps ferromagnetic states onto their antiferromagnetic counterparts.

The six ordered phases in this model can be summarized as follows: (i) \textit{$x$--$y$ N\'eel}: an in-plane N\'eel antiferromagnet with four variants (spins along $\pm\hat{x}$ or $\pm\hat{y}$). (ii) \textit{Stripe $\parallel$}: an in-plane stripe state with spins collinear with the stripe (ordering-wave-vector) direction. (iii) \textit{$z$ FM}: a uniform ferromagnet polarized along $\pm\hat{z}$. (iv) \textit{$x$--$y$ FM}: an in-plane ferromagnet with four variants (spins along $\pm\hat{x}$ or $\pm\hat{y}$). (v) \textit{Stripe $\perp$}: an in-plane stripe state analogous to Stripe $\parallel$, but with spins perpendicular to the stripe direction. (vi) \textit{$z$ AFM}: a N\'eel antiferromagnet with staggered spins polarized along $\pm\hat{z}$. The two $z$-polarized phases are Ising-like and have a twofold degeneracy. For the four in-plane phases, the low-temperature ordered states are selected by thermal order-by-disorder from an accidentally degenerate manifold~\cite{PhysRevB.110.104426}; the selected states break the coupled spin--lattice $C_{4}$ symmetry and an in-plane spin-reversal ($Z_{2}$) symmetry, resulting in a fourfold degeneracy. Finally, Klein duality~\cite{PhysRevB.110.104426} provides an explicit parameter mapping via a four-sublattice transformation, relating the phases pairwise: (i) to (ii), (iii) to (vi), and (iv) to (v).

We perform large-scale Monte Carlo simulations combining standard Metropolis updates with over-relaxation steps for efficient sampling.
We run multiple independent Markov chains in parallel (typically 56--112) to improve statistics for standard observables. For each parameter set, each chain is equilibrated for at least $10^{6}$ Monte Carlo sweeps (MCS) and then continued for at least $10^{6}$ MCS for measurements. For histogram-based analyses, we instead employ a single long Markov chain and further increase the statistics such that the total number of accumulated samples reaches at least $10^{8}$ after thermalization, yielding well-resolved probability distributions. Convergence is checked by confirming that the histograms (including peak positions and relative weights) are stable upon further increasing the sampling.
The simulations are performed on square lattices of size $L\times L$ with periodic boundary conditions, with $L$ up to $160$, ensuring reliable finite-size scaling analyses.

Khatua \textit{et al.} characterized the low-temperature phases using a composite in-plane order parameter capturing both magnetic and lattice aspects of ordering~\cite{PhysRevB.110.104426}. Here we disentangle the corresponding symmetry breakings to resolve the nature of the thermal transitions. 
Specifically, in-plane ($xy$) magnetic order breaks the joint inversion $(S^x,S^y)\!\to\!-(S^x,S^y)$ together with the spin–lattice $C_4$ symmetry, producing coupled magnetic–nematic order, 
whereas the $z$-polarized phases, which break only $S^z\!\to\!-S^z$, exhibit a conventional Ising-type transition.  
To characterize these ordered phases and their critical behavior, 
we monitor both nematic and magnetic order parameters, along with their Binder cumulants and long-distance correlation functions.

The transition associated with the breaking of the spin--lattice $C_{4}$ symmetry is characterized by the nematic order parameter~\cite{PhysRevB.109.195131,PhysRevE.81.066702},
\begin{equation}
N=  \frac{1}{L^2} \left|\sum_i{ (S_{i}^{x}S_{i+\hat{x}}^{x}-S_{i}^{y}S_{i+\hat{y}}^{y}) }\right|,
\end{equation}
which captures the spontaneous selection of a preferred lattice direction. 
To analyze its critical behavior, we compute the Binder cumulant
\begin{equation}
U_N = \frac{1}{2}(3 - \frac{\langle N^4 \rangle}{\langle N^2 \rangle^2}),
\end{equation}
and the corresponding two-point correlation function
\begin{equation}
C_N(\mathbf{r}) = \Big\langle \frac{1}{L^2} \sum_{i} N_i N_{i+\mathbf{r}} \Big\rangle,
\end{equation}
where $N_i = S_i^x S_{i+\hat{x}}^x - S_i^y S_{i+\hat{y}}^y$ and we evaluate it at the maximum separation as $C_N = C_N(L/2,L/2)$.

We next consider the observables characterizing the magnetic phase transitions, associated with the spontaneous breaking of spin-inversion symmetries. 
For each spin component $\alpha \in \{x,y,z\}$, we define the magnetization at its ordering wavevector $\mathbf{Q}_\alpha$ as
\begin{equation}
M_{\alpha} = \frac{1}{L^2}\sum_{i} S_i^{\alpha}\, e^{\,i\,\mathbf{Q}_{\alpha}\!\cdot \mathbf{r}_i}.
\label{eq:m_alpha_Q}
\end{equation}
For the four in-plane ordered phases (x-y N\'eel, Stripe $\parallel$, x-y FM, and Stripe $\perp$), the relevant magnetic order parameter combines the $x$ and $y$ spin components,
\begin{equation}
M_{xy} = \sqrt{M_x^2 + M_y^2},
\label{eq:Mxy_def}
\end{equation}
with the corresponding ordering wavevectors
\begin{equation}
\begin{aligned}
\text{x-y } \text{N\'eel}: \quad & (\mathbf{Q}_x,\mathbf{Q}_y)=((\pi,\pi),(\pi,\pi)),\\
\text{ Stripe }{\parallel}: \quad & (\mathbf{Q}_x,\mathbf{Q}_y)=((\pi,0),(0,\pi)),\\
\text{x-y } \text{FM}: \quad & (\mathbf{Q}_x,\mathbf{Q}_y)=((0,0),(0,0)),\\
\text{ Stripe }{\perp}: \quad & (\mathbf{Q}_x,\mathbf{Q}_y)=((0,\pi),(\pi,0)).
\end{aligned}
\label{eq:Qxy_table}
\end{equation}
For the out-of-plane ordered phases (z FM and z AFM), which break only the spin-inversion symmetry $S^z\!\to\!-S^z$, we use $|M_z|$
as the order parameter, with ordering vectors $\mathbf{Q}_z=(0,0)$ and $(\pi,\pi)$, respectively. 
The Binder cumulant for the in-plane magnetic order parameter is defined as
\begin{equation}
U_{M} = 2 - \frac{\langle M_{xy}^4\rangle}{\langle M_{xy}^2\rangle^2},
\end{equation}
while that for the $z$-polarized phases is defined analogously (see Appendix~\ref{app:Ising}).
We also compute the spin--spin correlation function
\begin{equation}
C_M(\mathbf{r}) = \Big\langle \frac{1}{L^2}\sum_{i} \mathbf{S}_i \cdot \mathbf{S}_{i+\mathbf{r}} \Big\rangle,
\label{eq:CS}
\end{equation}
which is evaluated at the maximum separation $C_M = C_M(L/2,L/2)$.

\section{Results \label{sec:results}}

In this section, we present numerical results for the finite-temperature phase transitions of the classical Heisenberg--compass model.
We first focus on the transitions into the in-plane ordered phases.
Sec.~III~A analyzes the regime of continuous transitions, extracting the varying correlation-length exponent $\nu$ and the anomalous dimensions $\eta_N$ and $\eta_M$ for the nematic and magnetic orders. 
Sec.~III~B identifies the four-state Potts points that terminate the Ashkin--Teller lines, and Sec.~III~C characterizes the adjacent first-order boundaries emerging beyond them. Finally, Sec.~III~D examines the $z$-polarized phases, whose thermal transitions belong to the conventional two-dimensional Ising universality class.

\subsection{Ashkin--Teller criticality}

\begin{figure}
\includegraphics[height=7.0cm,width=8.6cm]{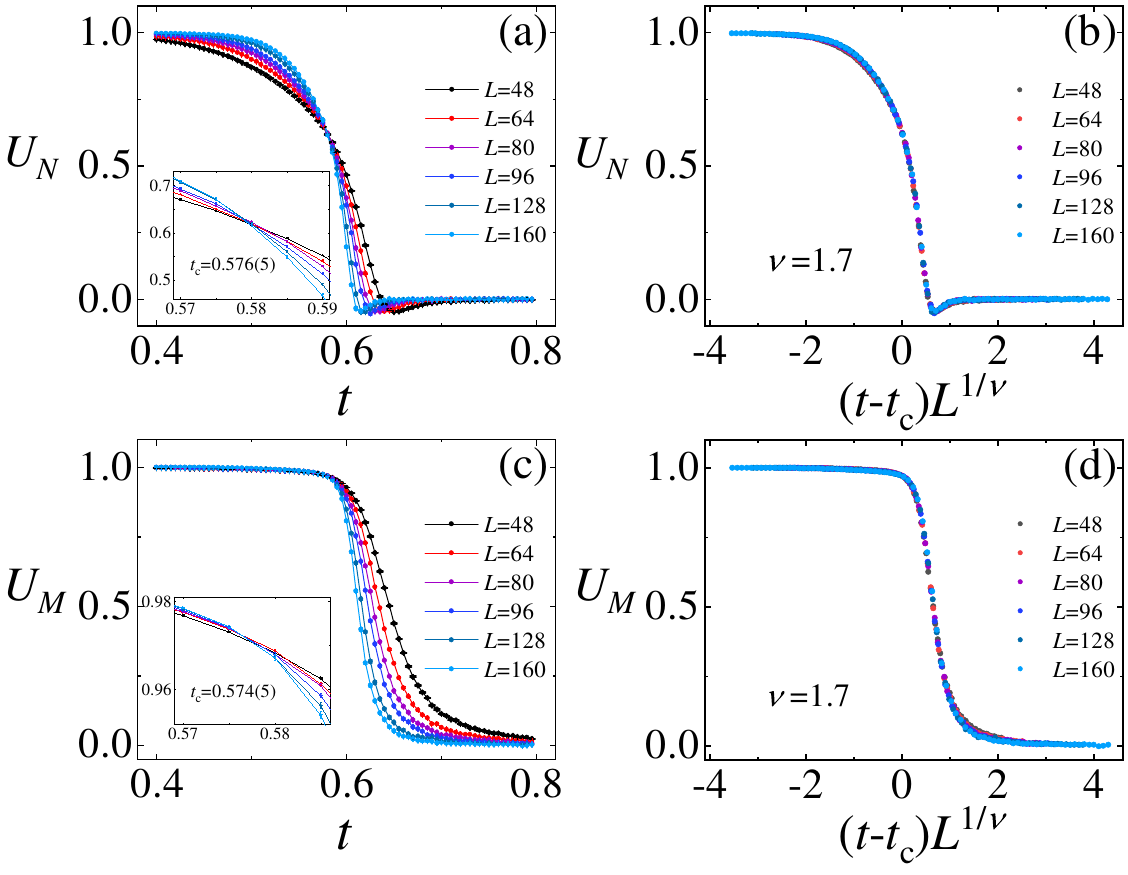}
\caption{
Finite-size scaling of the Binder cumulants at $\phi = 0.35\pi$ across the transition into the x-y N\'eel phase. 
(a,c) Nematic ($U_N$) and magnetic ($U_M$) cumulants exhibit crossings at a common critical temperature $t_c$ (zoomed-in insets highlight the crossing regions). 
(b,d) Corresponding scaling collapses according to $U = F[(t - t_c)L^{1/\nu}]$ give a consistent exponent $\nu = 1.70(5)$.
A shallow negative dip in $U_N$ is attributed to strong fluctuations of the bond-based nematic order parameter, which can produce nonmonotonic Binder behavior even for a continuous transition, unlike the pronounced negative minimum that deepens systematically with $L$ in the first-order regime discussed later.} 

\label{fig:2}
\end{figure}
We begin by analyzing the finite-temperature transition into the x-y N\'eel phase. 
Figure~\ref{fig:2} shows the nematic and magnetic Binder cumulants, $U_N$ and $U_M$, at $\phi = 0.35\pi$. 
Instead of employing the $(L,2L)$ crossing analysis used in Ref.~\cite{PhysRevB.109.195131}, 
we determine the correlation-length exponent~$\nu$ directly from finite-size scaling collapses according to 
$U = F[(t - t_c)L^{1/\nu}]$. 
The analysis spans a broad range of system sizes, $L = 48$--$160$, 
providing a sufficiently wide scaling window to capture the critical behavior. 
The Binder cumulants for different~$L$ exhibit consistent crossings and excellent scaling collapse, 
indicating that the smallest lattice ($L=48$) is already large enough to resolve the correct scaling behavior of this transition.
The common critical temperature obtained from $U_N$ and $U_M$ demonstrates that the nematic and magnetic orderings emerge simultaneously, 
signaling a single transition that couples the spin--lattice $C_4$ and spin-inversion symmetries without an intermediate nematic phase.
The correlation-length exponent $\nu = 1.70(5)$, extracted independently from both channels, confirms this concurrent symmetry breaking and indicates
nontrivial critical behavior distinct from conventional universality classes.

\begin{figure}
\includegraphics[height=3.6cm,width=8.8cm]{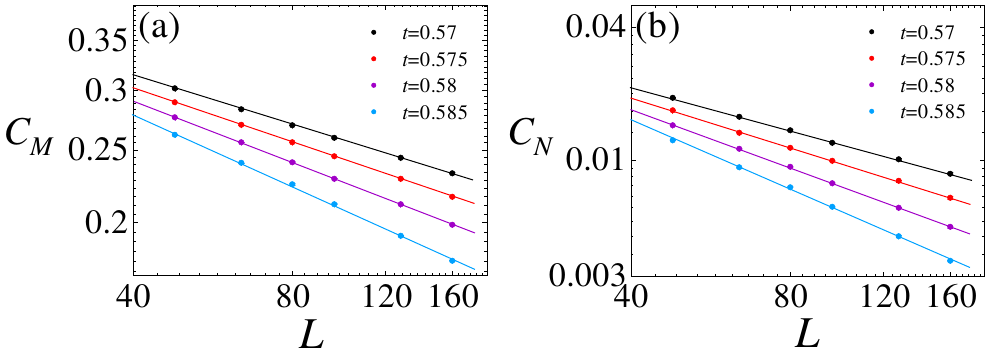}
\caption{
Finite-size scaling of the longest-distance correlations at $\phi = 0.35\pi$. 
(a) Magnetic anomalous dimension $\eta_M$ extracted from $C_M$ for system sizes $L = 48$–$160$, yielding $\eta_M = 0.214(2),\,0.240(3),\,0.273(3),\,0.321(6)$ at $t = 0.57,\,0.575,\,0.58,\,0.585$, respectively. 
(b) Nematic anomalous dimension $\eta_N$ extracted from $C_N$ over the same system sizes, giving $\eta_N = 0.64(1),\,0.74(2),\,0.88(2),\,1.06(4)$. 
}
\label{fig:3}
\end{figure}

The corresponding anomalous dimensions are extracted from the longest-distance spin and nematic correlations.
At criticality the magnetic correlation function decays algebraically as $C_M(r)\sim r^{-(d-2+\eta_M)}$~\cite{Cardy1996}, which in $d=2$ reduces to $C_M(r)\sim r^{-\eta_M}$; evaluating it at the maximum separation gives $C_M(L/2,L/2)\sim L^{-\eta_M}$, which we use to extract $\eta_M$~\cite{PhysRevB.109.195131}. Analogously, the nematic correlator obeys $C_N(L/2,L/2)\sim L^{-\eta_N}$~\cite{PhysRevB.109.195131}.
At $\phi = 0.35\pi$, the fitted values vary systematically with temperature near the transition, yielding 
$\eta_M = 0.214(2)$, $0.240(3)$, $0.273(3)$, and $0.321(6)$ at 
$t = 0.57$, $0.575$, $0.58$, and $0.585$, respectively (fitting range $L = 48$--$160$). 
Only at the critical point does the fitted exponent converge to its universal value: 
the AT prediction $\eta_M = 1/4$ is realized near $t \simeq 0.575$, 
in excellent agreement with the critical temperature independently determined from the Binder-cumulant analysis in Fig.~\ref{fig:2}. 
This value follows from identifying the spin field $\mathbf{S}(\mathbf{r})$—odd under spin inversion—with the Ising spin field in the Ashkin--Teller theory. The scaling dimension of this field is fixed at $x_h = 1/8$ along the AT line, directly implying $\eta_M = 2 x_h = 1/4$\tr{~\cite{PhysRevB.109.195131,Nienhuis1987}}.

In contrast, the nematic order parameter probes the directional symmetry breaking between the $x$ and $y$ bonds and, in the Ashkin--Teller framework, corresponds to the “polarization” operator, which represents the product of the two coupled Ising variables.
The scaling dimension of this operator varies continuously along the AT line, 
leading to the analytic relation $\eta_N = 1 - 1/(2\nu)$ between the nematic anomalous dimension $\eta_N$ and the correlation-length exponent $\nu$\tr{~\cite{PhysRevB.109.195131,Nienhuis1987}}.   
From the nematic correlation $C_N$, we obtain $\eta_N = 0.64(1)$, $0.74(2)$, $0.88(2)$, and $1.06(4)$ at the same temperatures. 
Using our numerical estimate $\nu=1.70(5)$, this gives $\eta_N\simeq0.706$, also in agreement with the fitted value near $t_c = 0.575$. 
The simultaneous consistency of $\eta_M$ and $\eta_N$ with the AT scaling relations 
provides compelling evidence that the transition belongs to the AT universality class.

\begin{figure}
\includegraphics[height=5.0cm,width=6.8cm]{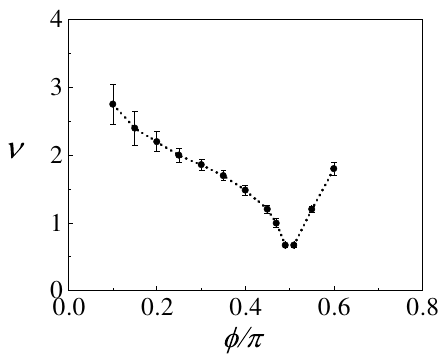}
\caption{Correlation-length exponent $\nu$ along the continuous phase boundaries of the x-y N\'eel and Stripe $\parallel$ phases, extracted from finite-size scaling of Binder cumulants. For $\phi>\pi$, the values coincide with those at $\phi-\pi$. 
The continuous variation of $\nu$ with $\phi$ is consistent with the Ashkin--Teller universality class.}
\label{fig:4}
\end{figure}

To further establish the universality class, we examine the evolution of the exponent $\nu$—together with the anomalous dimensions $\eta_N$ and $\eta_M$ (not shown)—as a function of $\phi$.
Figure~\ref{fig:4} reports the values of $\nu$ extracted along the continuous phase boundaries of the x-y N\'eel and Stripe $\parallel$ phases for
$\phi\in(0,\pi)$, revealing a clear systematic variation of $\nu$ across this range.
Additional $\phi$ values exhibit finite-size scaling behavior consistent with the representative cases shown in the text (e.g., $\phi=0.35\pi$); an extended dataset is available in Ref.~\cite{dataset}.
For $\phi>\pi$, the results map onto those at $\phi-\pi$, owing to a sublattice spin-flip transformation that relates the corresponding Hamiltonians.  
Altogether, these results establish that the in-plane continuous transitions belong to the Ashkin--Teller universality class.

\subsection{Four-state Potts point \label{sec:potts}}

\begin{figure}
\includegraphics[height=7.2cm,width=8.7cm]{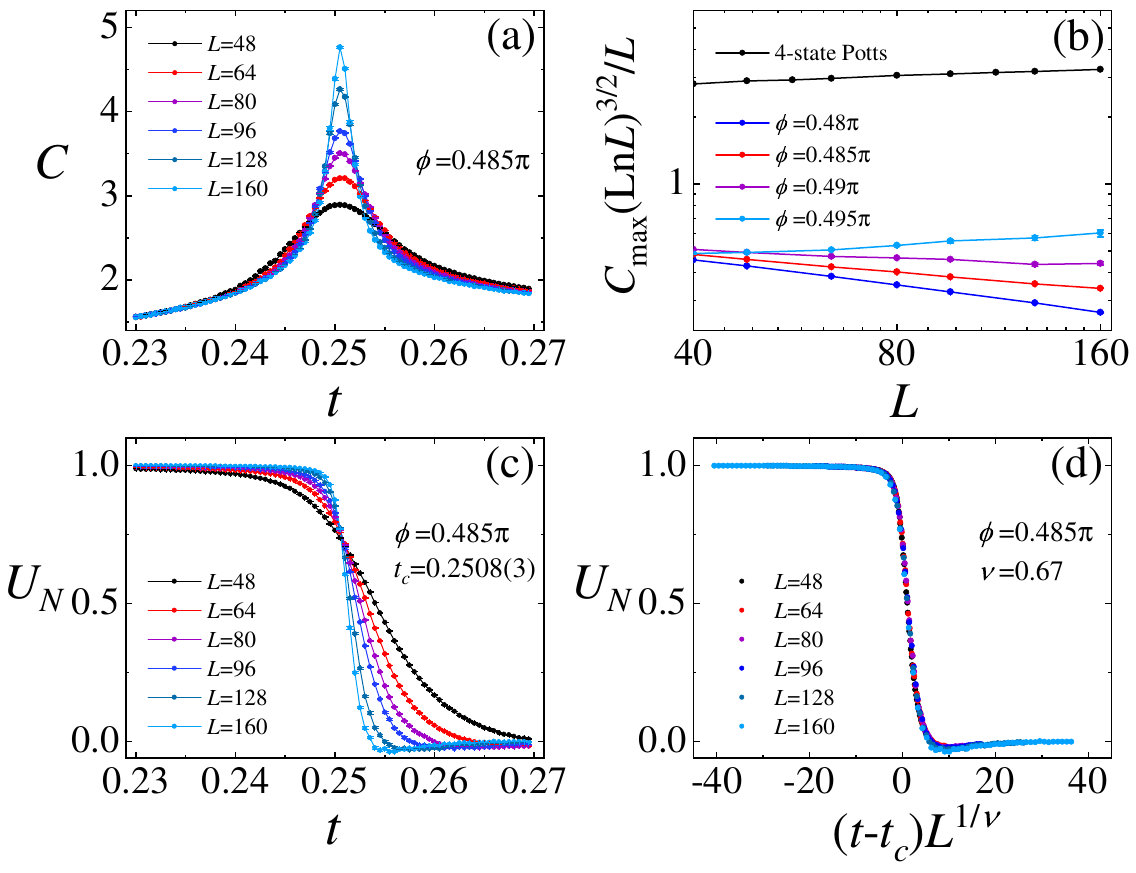}
\caption{
Identification of the four-state Potts critical point. 
(a) Specific heat $C(t)$ for several system sizes at $\phi = 0.485\pi$. 
(b) Scaling of the specific-heat peak $C_{\mathrm{max}}(L)$, plotted as $C_{\mathrm{max}}(\ln L)^{3/2}/L$, together with the same quantity for the four-state Potts model (black symbols). 
(c) Binder cumulant $U_N$ for several lattice sizes~$L$ at $\phi = 0.485\pi$, with the crossings yielding the critical temperature $t_c = 0.2508(3)$.
(d) Scaling collapse of $U_N$ yields a correlation-length exponent $\nu = 0.67(2)$, consistent with the four-state Potts value.
}
\label{fig:5}
\end{figure}

To complete the characterization of the Ashkin--Teller line, we next identify its termination at the four-state Potts critical point, beyond which the transition becomes first order~\cite{KADANOFF1979318}. 
This special point is located by analyzing the finite-size behavior of the specific heat and the Binder cumulants. 
Figures~\ref{fig:5}(a) and \ref{fig:5}(b) illustrate the thermodynamic behavior near the suspected Potts regime. 
In Fig.~\ref{fig:5}(a), the specific heat $C(t)$ develops a pronounced peak that sharpens and grows with increasing system size for fixed $\phi = 0.485\pi$, signaling a continuous phase transition and allowing an accurate extraction of its maximum value $C_{\mathrm{max}}(L)$.
In the four-state Potts model, $C_{\mathrm{max}}$ follows the characteristic form~\cite{PhysRevLett.44.837,Salas1997,PhysRevB.22.2560}
\begin{equation}
C_{\mathrm{max}} \propto L (\ln L)^{-3/2}.
\end{equation}
Accordingly, plotting $C_{\mathrm{max}}(\ln L)^{3/2}/L$ should yield a size-independent value precisely at the Potts point.  
As shown in Fig.~\ref{fig:5}(b), this behavior emerges within a narrow window $\phi \simeq 0.485\pi$--$0.49\pi$, 
thereby identifying the four-state Potts critical coupling at $\phi_{\mathrm{P}} \simeq 0.485\pi$--$0.49\pi$.

To confirm this identification, we perform an independent finite-size scaling analysis of the nematic Binder cumulant $U_N$ at $\phi = 0.485\pi$ [Figs.~\ref{fig:5}(c) and~\ref{fig:5}(d)]. 
The crossings of $U_N$ for different system sizes yield the critical temperature $t_c = 0.2508(3)$.  
A subsequent scaling collapse of $U_N$ according to $U_N = F[(t - t_c)L^{1/\nu}]$ gives the best-fit exponent $\nu = 0.67(2)$, 
in excellent agreement with the universal four-state Potts value $\nu = 2/3$.
Together with the specific-heat analysis, these results establish a four-state Potts critical point near $\phi \simeq 0.485\pi$--$0.49\pi$ that marks the termination of the Ashkin--Teller line.
The identification of the four-state Potts point in the thermal transition of the Stripe $\parallel$ phase follows an analogous procedure and is detailed in Appendix~\ref{app:phaseII}.

\subsection{First-order transition \label{sec:first}}

\begin{figure}
\includegraphics[height=7.2cm,width=8.8cm]{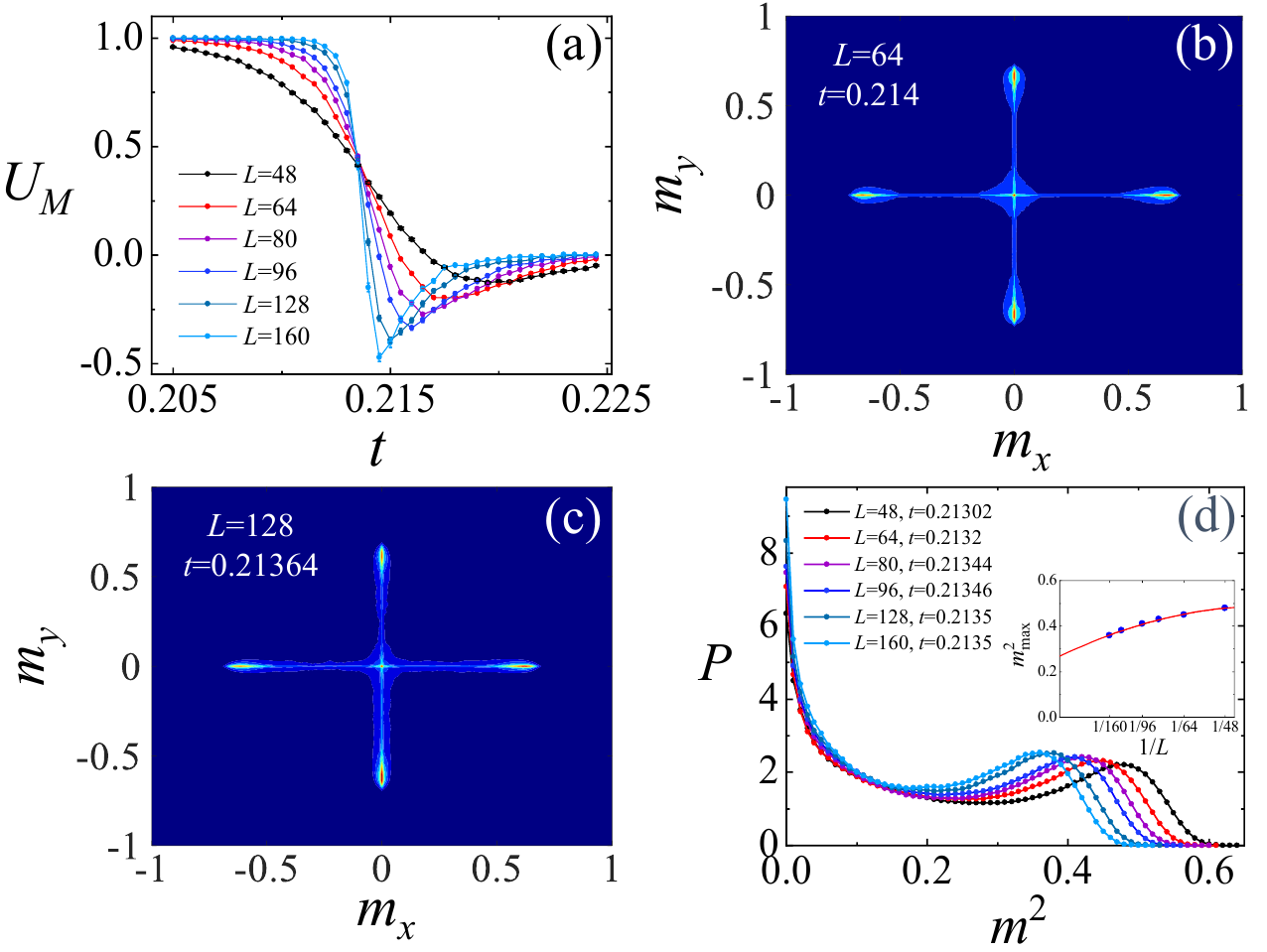}
\caption{
Evidence of a first-order transition at $\phi = 0.492\pi$. 
(a) Binder cumulant $U_M$ versus temperature for several system sizes. 
(b),(c) Two-dimensional histograms of the in-plane magnetization $\mathbf{m} = (m_x, m_y)$ for $L=64$, $t=0.214$ and $L=128$, $t=0.21364$, 
revealing clear phase coexistence. 
(d) Probability density histogram of $m^2$ for various $L$; inset: finite-size extrapolation of the high-$m^2$ peak $m^2_{\mathrm{max}}$ using a polynomial fit.
}
\label{fig:6}
\end{figure}

Having established the four-state Potts critical point, we now turn to the behavior of the system for $\phi > \phi_{\mathrm{P}}$.
In this regime, the nature of the thermal transition undergoes a qualitative change: the continuous transition gives way to a first-order transition.
We demonstrate this by analyzing the Binder cumulants and the histogram evidence of phase coexistence at $\phi = 0.492\pi$.

Figure~\ref{fig:6}(a) presents the Binder cumulant $U_M$ as a function of temperature for several lattice sizes. 
Unlike continuous transitions, where $U_M$ increases monotonically from $0$ in the disordered phase to values near $1$ in the ordered phase as the temperature is lowered, here it develops a pronounced negative dip.
The magnitude of this dip increases systematically with $L$, 
a well-known finite-size signature of a first-order transition~\cite{PhysRevB.30.1477,Vollmayr1993}. 
This conclusion is further supported by the probability distribution of the magnetization.
Figure~\ref{fig:6}(b,c) shows the two-dimensional histogram of the vector magnetization $\mathbf{m} = (m_x, m_y)$ obtained from Monte Carlo sampling
at temperatures near the negative dip. 
Two distinct features are visible: one centered near the origin, corresponding to the disordered phase characterized by vanishing magnetization, 
and another comprising four discrete clusters that represent the four in-plane ordered states. 
The coexistence of these two features provides direct evidence of a first-order transition.

Complementary information is provided by the probability density histogram of $m^2$, shown in Fig.~\ref{fig:6}(d). 
For each system size, $P(m^2)$ is measured at the temperature where the probabilities of the ordered and disordered phases are equal, at which the distribution exhibits a pronounced bimodal structure with two well-separated peaks.
The low-$m^2$ peak corresponds to the disordered phase, while the high-$m^2$ peak originates from the ordered phase. 
As the system size $L$ increases, the two peaks become sharper, reflecting the increasing stability of phase coexistence.
This stability is further quantified by tracking the finite-size evolution of the high-$m^2$ peak position, $m_{\mathrm{max}}^2$, extracted for each $L$ and extrapolated to the thermodynamic limit, as shown in the inset of Fig.~\ref{fig:6}(d).
The extrapolated value of $m_{\mathrm{max}}^2$ remains finite, indicating robust phase coexistence at the transition in the thermodynamic limit. 
Together, these results establish the four-state Potts point as the endpoint of the AT line, beyond which the transition is first order.
Moreover, similar analyses performed at couplings even closer to the compass limit confirm that the transition remains first order for all couplings between the Potts point and the pure compass point.
An analogous analysis for the Stripe $\parallel$ phase transition, presented in Appendix~\ref{app:phaseII}, likewise confirms its first-order character.

\subsection{Ising transitions in the $z$-polarized phases \label{sec:Ising}}

Finally, we examine the thermal transitions into the $z$-polarized ferromagnetic and antiferromagnetic phases. 
Both transitions break only the spin-inversion symmetry $S^z\!\to\!-S^z$ while preserving the spin–lattice $C_4$ symmetry, 
and are therefore expected to belong to the two-dimensional Ising universality class. 
As shown in Fig.~\ref{fig:S3} (see Appendix~\ref{app:Ising}), 
finite-size scaling of the Binder cumulants and the order parameters show excellent consistency with Ising critical behavior.
These results firmly establish the Ising character of the $z$-polarized transitions 
and complete the classification of all six ordered phases in the model.
We also note that Klein duality, combined with a sublattice spin-flip transformation, connects the four in-plane ordered phases and maps the z FM phase onto the z AFM phase.

\section{Summary \label{sec:discussion}}

Our study completes the characterization of the finite-temperature phase transitions of the Heisenberg--compass model, a problem that has remained unresolved despite earlier progress on its excitations~\cite{PhysRevLett.130.266702} and ground-state properties~\cite{PhysRevB.110.104426}. 
By mapping out the full thermal phase diagram and identifying the associated universality classes, we show that phases breaking both the spin–lattice $C_{4}$ symmetry and the in-plane spin-inversion symmetry realize Ashkin--Teller criticality with continuously varying exponents, with critical lines ending at four-state Potts points before turning first order.

This should be contrasted with the pure compass limit, where the model possesses a one-dimensional subsystem symmetry that forbids finite-temperature magnetic ordering. As a result, only nematic order develops, reducing the spin–lattice symmetry from $C_4$ to $C_2$. 
Introducing Heisenberg exchange explicitly removes this subsystem symmetry, thereby allowing the in-plane spin-inversion $Z_2$ and the nematic $Z_2$ to break simultaneously. 
The emergence of magnetic order also eliminates the remaining $C_2$ symmetry and fully breaks $C_4$, giving rise to the Ashkin–Teller criticality.
A direct comparison with the generic compass model (gCM) of Ref.~\cite{PhysRevB.109.195131} further elucidates this mechanism.
While the XY limit of the gCM exhibits a Kosterlitz–Thouless transition and the Heisenberg limit of our model shows only a thermal crossover, both systems display Ashkin–Teller criticality throughout the regime where $C_4$ symmetry is broken.
This indicates that the universal behavior is ultimately governed by the discrete spin–lattice $C_4$ symmetry.

Theoretically, attention should turn to the quantum Heisenberg–compass model, exploring how quantum fluctuations reshape its phase diagram, including the robustness of the universality-class structure established here and the possible renormalization of the tunable critical exponent $\nu$. On the experimental side, the critical signatures and universal scaling relations identified in our analysis provide concrete guidance for candidate square-lattice magnets featuring compass-like anisotropic exchanges~\cite{PhysRevLett.110.117207,PhysRevX.4.021051,PhysRevB.94.161118,annurev:/content/journals/10.1146/annurev-conmatphys-031218-013113,PhysRevB.83.155118}.

\begin{acknowledgments}
We thank Rong Yu, Changle Liu, and Fan Zhang for valuable discussions. This work was supported by the National Natural Science Foundation of China (Grant Nos. 12404176 and 12564022).
\end{acknowledgments}

\section*{Data Availability}
The data that support the findings of this study are openly available~\cite{dataset}.

\appendix 

\section{Nematic and magnetic order parameters at $t = 0.05$}\label{app:order}
\begin{figure}[!ht]
\centering
\includegraphics[height=7.08cm,width=7.5cm]{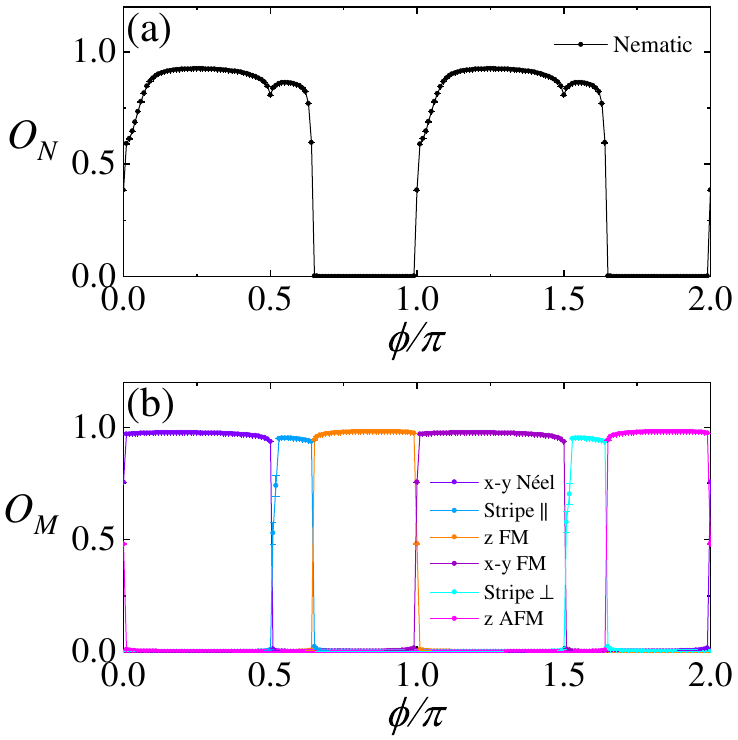}
\caption{Nematic (a) and magnetic (b) order parameters as functions of $\phi$ at $t = 0.05$, obtained for system size $L=64$.}
\label{fig:S1}
\end{figure}

Figure~\ref{fig:S1} shows the results for the nematic and magnetic order parameters at $t = 0.05$.

\section{Finite-temperature phase transition into the Stripe $\parallel$ phase}\label{app:phaseII}
\begin{figure}[!ht]
\centering
\includegraphics[height=6.9cm,width=8.6cm]{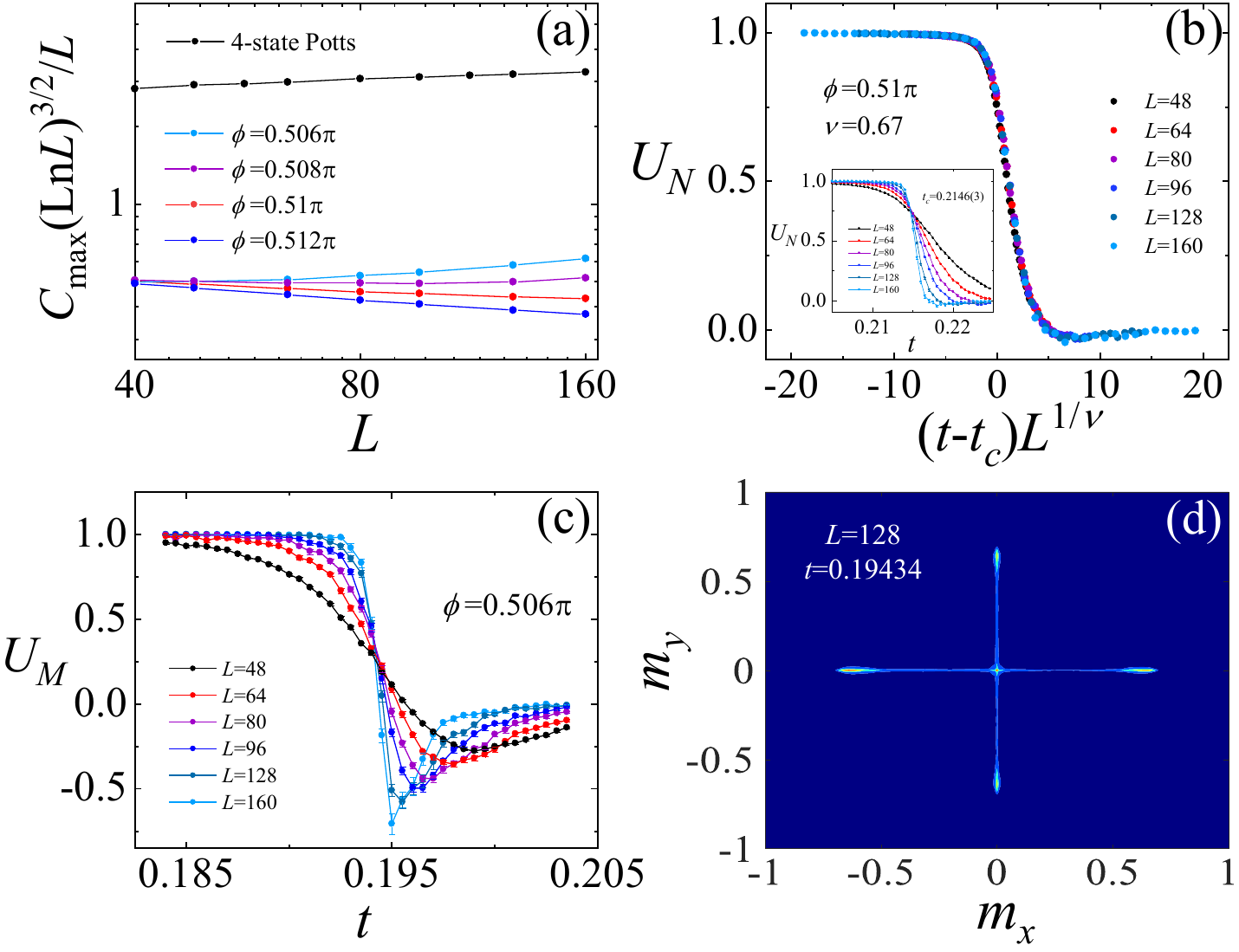}
\caption{(a) Scaling of the specific-heat peak $C_{\mathrm{max}}(L)$ identifies the four-state Potts coupling $\phi_{\mathrm{P}} \simeq 0.508\pi$--$0.51\pi$. 
(b) Scaling collapse of $U_N$ at $\phi = 0.51\pi$ yields $\nu = 0.67(2)$, consistent with the Potts value. 
(c) Binder cumulant $U_M$ versus temperature at $\phi = 0.506\pi$, showing a negative dip that deepens with $L$, characteristic of a first-order transition. 
(d) Two-dimensional histogram of the in-plane magnetization $\mathbf{m} = (m_x, m_y)$ for $L=128$, $t=0.19434$ at $\phi = 0.506\pi$.}
\label{fig:S2}
\end{figure}

Figure~\ref{fig:S2} shows the results for the finite-temperature phase transition into the Stripe $\parallel$ phase.

\section{Ising transition into the z FM phase}\label{app:Ising}
\begin{figure}[!ht]
\centering
\includegraphics[height=7.0cm,width=5.2cm]{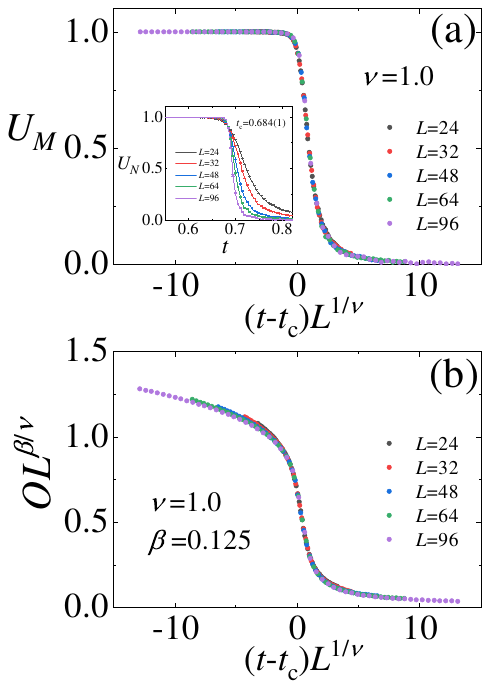}
\caption{Evidence for a two-dimensional Ising transition into the z FM phase at $\phi = 0.8\pi$.
(a,b) Finite-size scaling of the Binder cumulant and the order parameter, yielding critical exponents consistent with the two-dimensional Ising universality class.
The Binder cumulant for the $z$-polarized phases is defined as $U_M = \frac{1}{2}\left(3 - \frac{\langle M_z^4 \rangle}{\langle M_z^2 \rangle^2}\right)$.
Scaling collapses use $U_M = F_U[(t - t_c)L^{1/\nu}]$ and $O = L^{-\beta/\nu}F_O[(t - t_c)L^{1/\nu}]$.}
\label{fig:S3}
\end{figure}

Figure~\ref{fig:S3} shows the results for the Ising transition at $\phi = 0.8\pi$.

\newpage
\bibliography{heisenberg_compass.bib}

\end{document}